\documentclass[journal]{IEEEtran}
\IEEEoverridecommandlockouts
\usepackage{cite}
\usepackage{amsmath,amssymb,amsfonts}
\usepackage{algorithmicx}
\usepackage{graphicx}
\usepackage{textcomp}
\usepackage{xcolor}
\usepackage{amsmath}
\usepackage{amsfonts}
\usepackage{graphicx}
\usepackage[colorinlistoftodos]{todonotes}
\usepackage{algorithmicx}
\usepackage[ruled]{algorithm}
\usepackage{algpseudocode}
\usepackage{comment}
\usepackage[letterpaper, portrait, margin=0.5in]{geometry}
\def\BibTeX{{\rm B\kern-.05em{\sc i\kern-.025em b}\kern-.08em
T\kern-.1667em\lower.7ex\hbox{E}\kern-.125emX}}

\usepackage{enumerate}
\usepackage{enumitem}
\usepackage{bm}
\usepackage{amssymb}
\usepackage{amsmath}
\usepackage{amsthm}

\newtheorem{remark}{Remark}

\begin{document}

\title{Fast Quantum Concentration via Grover's Search}

\author{\IEEEauthorblockN{Cem M Unsal}\\
\IEEEauthorblockA{\textit{Department of Mathematics} \\
cem@umd.edu\\}
\and
\vspace{5pt}
\IEEEauthorblockN{A Yavuz Oruc}\\
\IEEEauthorblockA{\textit{Department of Electrical and\\
Computer Engineering} \\
yavuz@umd.edu}

\vspace{10pt}
\textit{University of Maryland, College Park, Maryland, 20742}

\vspace{-25pt}
}

\title{Faster Quantum Concentration via Grover's Search}

\maketitle

\begin{abstract}
We present quantum algorithms for routing concentration assignments on full capacity fat-and-slim concentrators, bounded fat-and-slim concentrators, and regular fat-and-slim concentrators.
Classically, the concentration assignment takes $O(n)$ time on all these concentrators, where $n$ is the number of inputs.
Powered by Grover's quantum search algorithm, our algorithms take $O(\sqrt{nc}\ln{c})$ time, where $c$ is the capacity of the concentrator. Thus, our quantum algorithms are asymptotically faster than their  classical  counterparts, when $c\ln^2{c}=o(n)$.
In general, $c = n^\mu,$ satisfies $c\ln^2{c}=o(n),$ implying a time complexity of $O(n^{0.5(1+ \mu )} \ln n),$ for any $\mu, 0 < \mu < 1.$

\end{abstract}

\noindent\vspace{-15pt}


\vspace{-14pt}
\section{Introduction}

\vspace{-3pt}\noindent
The promise of quantum computing in speeding up computations continues to attract research into exploring quantum algorithms for problems that arise in computer science, mathematics, and other scientific fields of study beyond searching and factorization.
Indeed, several quantum algorithms and circuits have been reported for a wide range of classical problems extending from algebraic computations to pattern matching and many problems in graph theory.
See for example~\cite{childs2010quantum} and \cite{mosca2008quantum} for a survey of quantum algorithms for Abelian and non-Abelian discrete Fourier transform, hidden subgroup problem, computing discrete logarithms, and several other problems in number theory, cryptography, and group theory.
Another article by Montanaro provides an overview of quantum algorithms, and in particular surveys the complexity of quantum searching and optimization algorithms~\cite{montanaro2016quantum}.
In graph theory, Grover's search algorithm and quantum walk techniques have been used to solve matching and network flow problems~\cite{AS06,ambainis2007quantum,Dor09} and graph traversals~\cite{childs2003exponential}.

In this paper, we focus our attention on a different direction, namely the application of quantum computing to developing fast quantum algorithms to realize connection requests in concentrators.
Concentration is a fundamental operation in data processing and is closely related to matchings in graph theory. Thus, we think that developing fast quantum algorithms for realizing connection requests in concentrators will likely have a significant impact on routing in packet switching networks.
Loosely speaking, concentration is a one-to-one mapping between two sets of objects, called inputs and outputs, where only inputs can be specified, and it is assumed that there are fewer outputs than inputs.
Formally, a graph with a set of vertices, representing inputs and another set of vertices, representing outputs, in which a one-to-one mapping exists between every subset of inputs and some subset of outputs over a set of non-overlapping paths is called a concentrator.
Pinsker established that concentrators with $n$ inputs, $m$ outputs, and at most $29n$ edges exist for all $m \le n$ in~\cite{pinsker73}.
Since Pinsker's seminal result, quite a few concentrator designs have been reported in the literature, see for example~\cite{pippenger77,chu78, bassalygo81,nakamuraMasson82,chienOruc94,orucHuang96, gunduzhanOruc97, GO98,ratanOruc2003, ratanOruc2010}.
The earlier designs in~\cite{pippenger77,chu78, bassalygo81} provide non-explicit solutions and can be viewed as upper bound results.
The first explicit concentrator was reported in \cite{nakamuraMasson82} using a binomial sparse crossbar design with $O(n^{1.5})$ edges.
The design given in \cite{chienOruc94} falls outside the focus of our work as it is based on binary comparators and sorters.
In this paper, we are concerned with designing classical and quantum algorithms for a particular set of concentrators that are referred to as  fat-and-slim sparse crossbars. These were introduced in~\cite{orucHuang96} and developed in~\cite{gunduzhanOruc97, GO98,ratanOruc2003, ratanOruc2010}. We note that concentrators are further refined to study a family of bipartite graphs, widely-known as expanders$\!$\cite{margulis73,gabberGalil78,Alon81,tanner84,jimboMaruoka85}. Expanders provide bounded capacity concentration with $O(n)$ edges. Our work on how to route concentration assignments in such graphs using quantum algorithms will be deferred to another place. In this paper, we give classical algorithms to realize concentration assignments in sparse crossbar concentrators and use Grover's search to transform them to quantum algorithms, decreasing their time complexity.
The rest of the paper is organized as follows. The next section provides the preliminary concepts needed to describe our results.
In Section \ref{section:models}, we state our assumptions of classical and quantum models of computation.
Sections \ref{section:classicalRouting} and \ref{section:quantumRouting} present our main results.
The paper is concluded in Section \ref{section:conclusion} with a discussion of our results and possible directions for future research.

\vspace{-10pt}
\section{Sparse Crossbar Concentrators}
\label{section:concentrators}

An $(n,m,c)$-concentrator is a graph with a set of $n$ vertices, called inputs, a set of $m$ vertices that are disjoint from the set of inputs, called outputs, and in which a $k$-matching, i.e., a set of $k$ vertex-disjoint paths exists between every $k$ inputs and some $k$ outputs, $1\le k\le c\le m$.
Such a concentrator is referred to as a bipartite or sparse crossbar concentrator if all paths are of length one, i.e., each path consists of two vertices and an edge that is represented by a crosspoint in the fat-and-slim crossbar model\cite{orucHuang96}.
If $c < m$ then an $(n,m,c)$-concentrator is said to have bounded capacity\footnote{The maximum number of inputs that can be concentrated in unit time independent of the location of the inputs \cite{fins}.}, and if $c =m,$ it is said to have full capacity.
Henceforth, we will refer to the latter as an $(n,m)$-concentrator.
A concentrator is called explicit if all of its edges (crosspoints) are specified, inexplicit if all of its edges exist with a non-vanishing probability other than 1, and semi-explicit if only a proper subset of its edges is explicitly specified.

\begin{figure}
\centerline{\includegraphics[width=\linewidth]{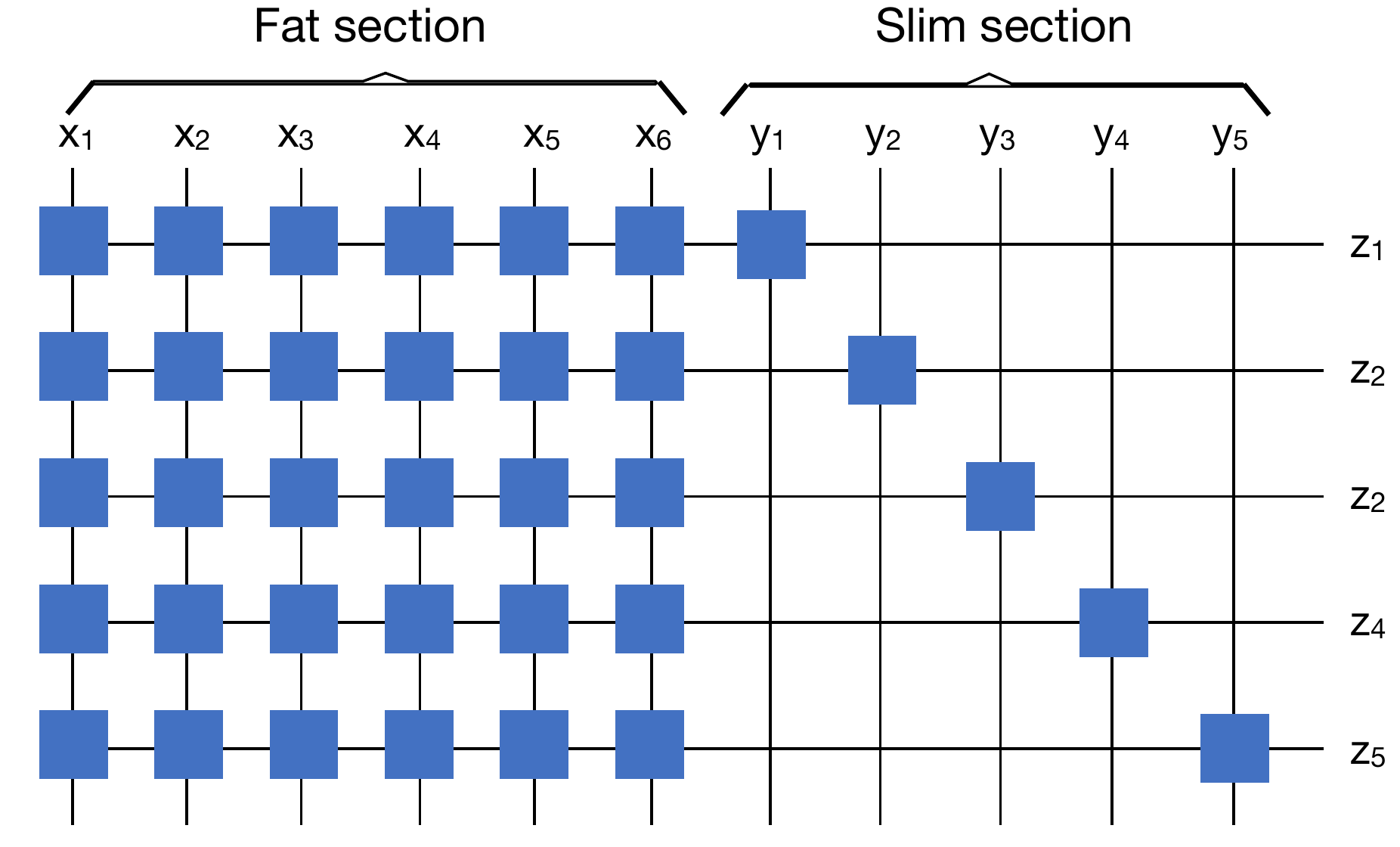}}
\vspace{-10pt}
\caption{An (11, 5) full-capacity, fat-and-slim concentrator.}
\label{fig1}
\end{figure}

Numerous explicit, semi-explicit, and inexplicit concentrators have been described in the literature as mentioned in the introduction.
In this paper, we focus on two explicit concentrator designs introduced in~\cite{orucHuang96} and one introduced in \cite{GO98}.
The first among these is a bipartite concentrator, called a full capacity fat-and-slim crossbar in which inputs are divided into two subsets of size $n-m$ and $m$.
Each of the $n-m$ inputs is connected to all the outputs, and forms the fat-part of the concentrator, whereas each of the $m$ inputs is connected to exactly one output and forms its slim part.
These concentrators are often diagrammed using a sparse crossbar representation as shown in Figure~\ref{fig1}.
As described, the left-hand side forms the fat part, and the diagonal line on right hand side forms the slim part of the concentrator.
Adding all the crosspoints (edges) together, we find that an fat-and-slim $(n,m)$-concentrator consists of $(n-m)m + m = (n-m+1)m$ crosspoints, and it was established in~\cite{orucHuang96} that  every bipartite $(n,m)$-concentrator must have at least $(n-m+1)m$ crosspoints. Therefore, the full-capacity fat-and-slim concentrator is minimal with respect to the number of crosspoints.

\begin{figure}
\vspace{-11pt}
\centerline{\includegraphics[scale=0.55]{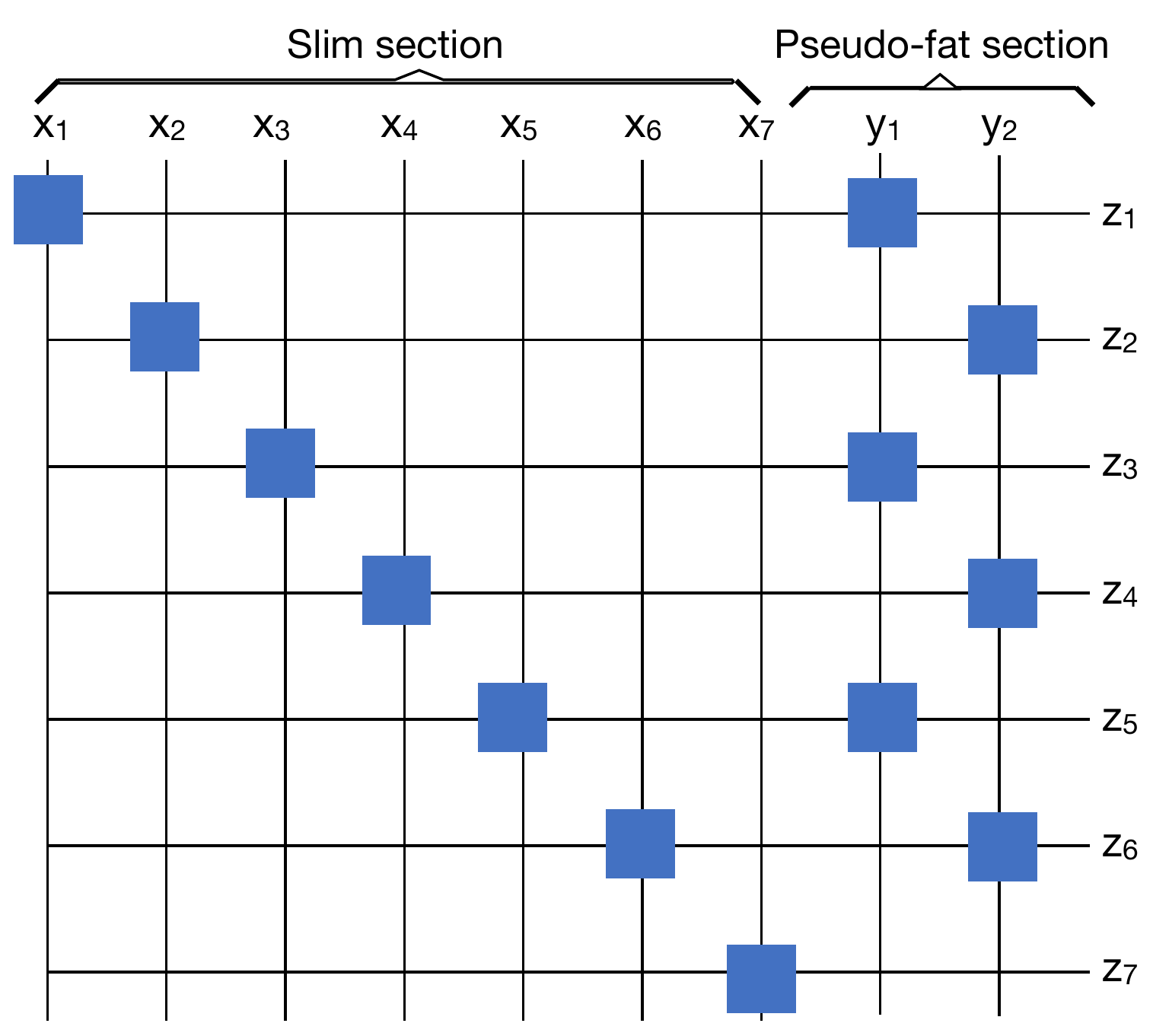}}
\caption{A bounded capacity fat-and-slim $(9,7,2)$-concentrator, which is also a $(9,7,3)$-concentrator by our extension of its capacity.
}
\vspace{-4pt}
\label{fig2}
\end{figure}

The second concentrator is also a sparse crossbar concentrator similar in concept to a full-capacity fat-and-slim concentrator, but with a bounded capacity $c$ as illustrated in Figure~\ref{fig2} for $n = 9$, $m = 7 = 5$, $c = 2$.
In this construction, the left-hand side forms the slim-part, whereas the right hand side is added to provide a minimum capacity of $c$.
This is ensured by requiring $c\le m/c$ or $c\le \sqrt{m}$.
The key idea is to provide a sufficient number of crosspoints (edges) to each input in the right part so that, if it gets blocked by as many as $c-1$ inputs from the slim part, it can still find an idle output to match with.
Requiring $c\le m/c$ secures this as each input in the right part is connected to $\lfloor m/c\rfloor$ outputs.
The constriction also assumes that $n-m\le c$.
This assumption ensures that the main diagonal of crosspoints in the slim section on the left spans the entire set of outputs.
It was shown in~\cite{orucHuang96} that the crosspoint complexity of this construction remains with a factor of two of a lower bound of $\lfloor\frac{(n-c+1)m}{m-c+1}\rfloor$ crosspoints.

We refer the reader to~\cite{orucHuang96} for a more in-depth account of these two concentrator constructions, while we note that the capacity of the second sparse crossbar concentrator is actually larger than $c$ when $m/c  >  c.$ In fact, the capacity of this crossbar is $\lfloor m/c\rfloor$ as each input in the fact section is connected to at least $\lfloor m/c\rfloor$ outputs. We adjust our notation to highlight this observation by replacing $c$ for the width of the inputs in the fat-section by $q,$ and letting $n-m\le q\le m,$ and $c=\lfloor m/q\rfloor.$ Therefore, $c\le \sqrt{m}$ is no longer required as described in~\cite{orucHuang96}.
For the rest of this paper, this updated construction will be referred to as ``bounded capacity fat-and-slim concentrator''.
Thus, we revise the capacity of the fat-and-slim crossbar  in Figure \ref{fig2} to 3. 

\begin{figure}
\vspace{5pt}
\centerline{\includegraphics[width=\linewidth]{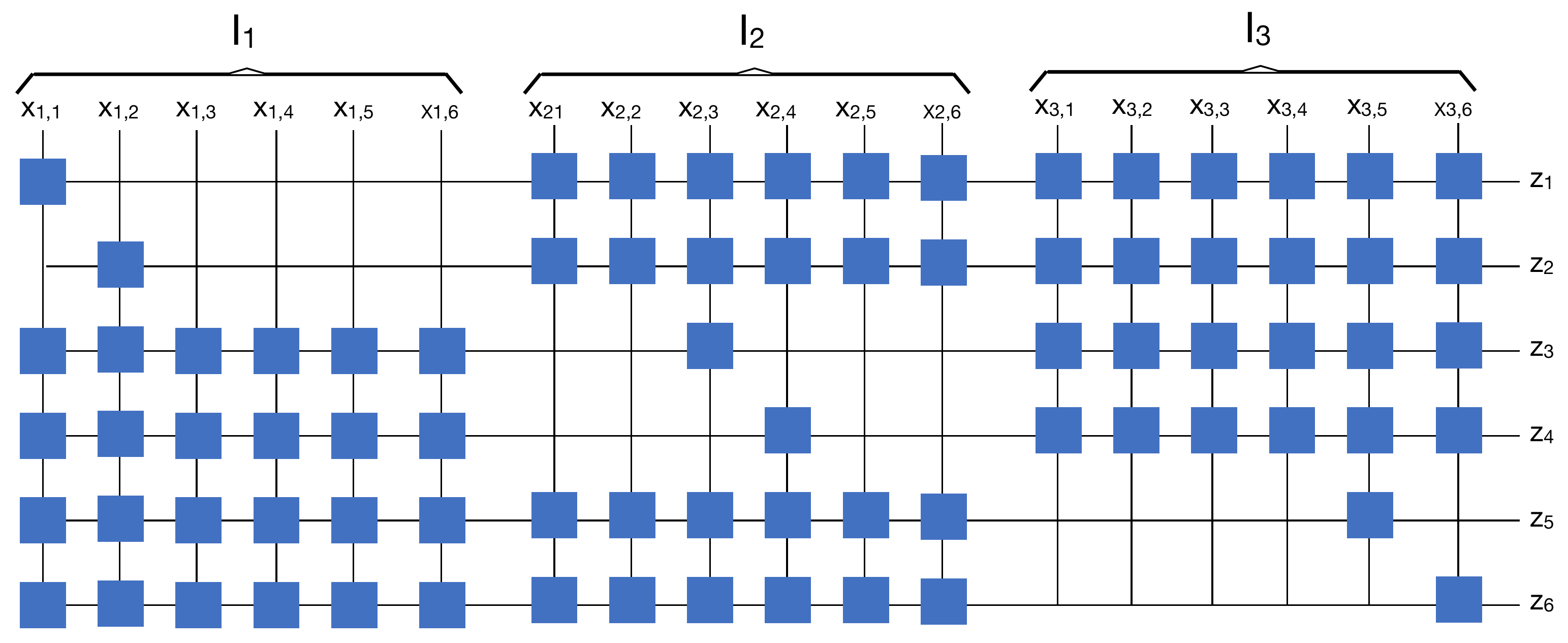}}
\vspace{-1pt}
\caption{A regular (18,6)-fat-and-slim concentrator.}
\label{fig3}
\end{figure}

In addition to these two concentrators, we will also present a quantum routing algorithm for a third sparse concentrator that has a more regular structure\cite{GO98}.
This concentrator uses $n = pm$ inputs and $m$ outputs, and it is derived from the full-capacity fat-and-slim concentrator.
Each output is connected to $n-m+1$ inputs as in the full-capacity fat-and-slim concentrator.
On the other hand, each input is connected to between $m-\lfloor m/p\rfloor$  and $m-\lfloor m/p\rfloor + 1$ outputs, making this $(pm,m)$-concentrator nearly regular in terms of its in-degree (fan-in) as well.
In this paper, we will consider the case when $p$ divides $m$.
Additionally, we only consider cases when $p\ge3$.
Figure~\ref{fig3} illustrates this construction for $p = 3,$ and $m = 6$.
It is seen that the out-degree of the construction falls between 4 and 5.
Effectively, this construction is obtained from the full-capacity fat-and-slim concentrator by (i) dividing $n = pm$ columns of crosspoints into $p$ sections,  (ii) chopping the diagonal of $m$ crosspoints in the slim part into $p$ groups of $m/p$ columns, and (iii) swapping each of those $m/p$ columns with an equal number of columns in one of the remaining $p-1$ sections on the left.
Using the notation in \cite{GO98}, we denote the sets of inputs in the $p$ sections by $I_j, 1\le j\le p,$ and let $I_j =\{x_{j,1},x_{j,2},\cdots, x_{j,m}\}$.
We further let $V_j = \cap_{k=1}^{m-1} N(x_{j,k}),$   $U_j = O\backslash V_j,$ and $W_j$ denote the set of inputs that are connected to the outputs in $U_j,$ where $N(x_{j,k})$ denotes the neighbor set of input $x_{j,k}, 1\le j\le p, 1\le k\le m$.
In Figure~\ref{fig3}, $V_1 = \{z_3, z_4, z_5, z_6\} , V_2 = \{z_1, z_2, z_5, z_6\},V_3 = \{z_1, z_2, z_3, z_4\}, U_1 = \{z_1,z_2\}, U_2 = \{z_3,z_4\}, U_3 = \{z_5, z_6\}, W_1 = \{x_{1,1}, x_{1,2}\}, W_2 = \{x_{2,3}, x_{2,4}\}, W_3 = \{x_{3,5}, x_{3,6}\}$.

The crux of this construction is a transformation theorem proved in~\cite{GO98}.
As we describe in the latter part of Section \ref{section:classicalRouting}, the construction of this concentrator makes  the design of a classical routing algorithm for  it more involved, but we  establish that the time complexity of such an algorithm remains $O(m)$ once active inputs are located.

\vspace{-15pt}
\section{Classical Versus Quantum Routing Model}
\label{section:models}

\vspace{-4pt}\noindent
The algorithms presented in this paper will be run on two models of computation: (\ref{subsection:classical}) classical model and (\ref{subsection:quantum}) quantum model.
It is important to highlight the differences between these two models to make a justifiable comparison of their execution times.

\vspace{-12pt}
\subsection{The Classical Routing Model}\label{subsection:classical}

\vspace{-2pt}\noindent
In the classical model, we assume that quantities of interest are represented using classical bits of 0 and 1.
For example, we use $\lg n$ bits and $\lg m$ bits to identify the inputs and outputs of an $(n,m)$-concentrator, respectively.
We further assume that the bits within the representation of each input and/or output can be processed in parallel.
For example, we can inspect all of the bits within a representation of any input (output) in parallel to determine if it is a particular input we seek, or we can compare the bits in representations of any two inputs (outputs) or a constant number of inputs (outputs) in parallel to see if they are the same in $O(1)$ time.
Such bit-level operations can generally be carried out by logic circuits that consist of elementary logic components such as OR, AND, XOR, XNOR, and NOT gates that we assume have $O(1)$ computation time, and have  a constant fan-in and fan-out, i.e., they have a constant number of inputs and outputs.
We note that two or a constant number of $\lg n$-bit numbers can be added, subtracted or compared using $O(\lg n)$ 2-input, 2-output logic gates in $O(\lg\lg n)$ time using a prefix-adder as described in~\cite{brent1982regular}.
This $O(\lg\lg n)$ time will be suppressed in our time complexity formulas, leaving us with $O(1)$ time.
It will also be assumed that $O(\lg n)$-bit operands can be written and read in and out of  a random access memory in $O(1)$ time.
We will be using an array named $in$ to specify if a given input is active.
We will assume that $in$ is an array that is stored in a random access memory with $O(1)$ access time.
It is possible to relax $O(1)$ access time complexity.
However, since such  a non-constant access time complexity would be a multiplicative coefficient in the time complexity of the bottleneck step of  both classical and quantum algorithms, we will omit this factor from our complexity calculations, effectively assume that the access time of  $in$ is $O(1).$
 
 \vspace{-18pt}
\subsection{The Quantum Routing Model}\label{subsection:quantum}

\vspace{-3pt}\noindent
In the quantum routing model, classical bits are replaced by quantum bits (qubits), and classical logic gates are replaced by those that represent unitary transformations. We assume that unitary gates have one  or  two inputs. All together, we allow $O(\lg n)$ qubits in the quantum routing model. This allows us to work with possibly up to $n$ states in parallel.
We further allow any combination of quantum gates to be used on these $O(\lg n)$ qubits, but with the restriction that each qubit can source only one quantum gate at a time.
This last restriction is a corollary of the no-cloning theorem\cite{buvzek1996quantum}.
This fanout restriction of one in the quantum routing model is not imposed on the bits in the classical routing model.
However, we still assume that the fanout of each input is $O(1)$ in the classical model as well.
Therefore, the two models can be viewed to be analogous, where any quantum operation on mutually exclusive $O(\lg n)$ qubits takes $O(1)$ time much the same way each operation on $O(\lg n)$ classical bits takes $O(1)$ time.
The difference lies in the amount of parallelism afforded by the two models: in the classical model, we assume that the parallelism is limited to operations on any given pattern of $O(\lg n)$ bits, whereas in the quantum routing model, the parallelism transcends any particular or fixed pattern as quantum operations are applied to the totality of the quantum state that includes all $O(n)$ binary patterns of $O(\lg n)$ qubits.
This vast amount of parallelism inherently present in quantum mechanical systems resulted in Shor's quantum prime number factorization algorithm\cite{shor1994algorithms}, which is exponentially faster than the best-known classical algorithm.
Another key quantum algorithm, Grover's quantum search\cite{Gro96} provides a speed-up of $O(\sqrt n)$ over a sequential algorithm to search an element in an unordered list of $n$ elements.
The latter algorithm will be used in Section~\ref{section:quantumRouting} to reduce the routing time of concentration assignment on fat-and-slim concentrators using quantum algorithms.
Before we describe these algorithms, we provide their classical analogs in the next section.

\vspace{-14pt}
\section{Classical Routing On Concentrators}
\label{section:classicalRouting}

\vspace{-4pt}\noindent
Routing an assignment on a sparse crossbar concentrator amounts to constructing a matching between any given subset of inputs and some subset of outputs of equal cardinality.
The particular topologies of the three concentrators given in Section~\ref{section:concentrators} guide the design of a routing algorithm for each concentrator as we describe next.

\vspace{-12pt}
\subsection{Full-Capacity Fat-Slim (F-S) Concentrator Routing}

\vspace{-2pt}\noindent
Our first algorithm, Algorithm~\ref{alg:classicalFull} is a restatement of the algorithm that was originally described in~\cite{guo96} for a full-capacity fat-and-slim concentrator. We recall from Section~\ref{section:concentrators} that the set of inputs is partitioned into two sets:
those in the fat section: $X = \{x_1,x_2,\cdots\!, x_{n-m}\}$ and those in the slim section: $Y = \{y_1,y_2,\cdots\!,y_{m}\}$.
The set of outputs of the concentrator is denoted by $Z=\{z_1,z_2,\cdots\!,z_{m}\}$.

\begin{algorithm}
\caption{Classical Full-Capacity F-S Concentration}
\label{alg:classicalFull}
\small
\begin{algorithmic}[1]
\vspace{2pt}
\Function{Classical Full FS Route}{$in,n,m$}\newline
//$in$: $n$-bit array that marks up to $m$ active inputs\newline 
//$n$: number of inputs.\newline 
//$m$: number of outputs.
\State $L \gets list()$

\For{$i \gets 1:m$} //slim section
\If {$in[m-n+i] == 1$}
\State$pair (y_i, z_i);$
\Else
\State $L.insert(z_i);$
\EndIf
\EndFor
\For {$i\gets 1:n-m$} //fat section
\If {$in[i]==1$}
\State $\{z=L.remove(); pair(x_i, z);\}$
\EndIf
\EndFor
\EndFunction
\end{algorithmic}
\end{algorithm}

\noindent
As we stated in the earlier section, the active inputs in a routing request, i.e., those to be concentrated are specified by `1' entries in an $n$-bit array, named {\em in}.
A routing request with $k$ active inputs is first completed to a request with $m$ inputs by combining the leftmost unused $m-k$ inputs in $in$ with the given $k$ active inputs.
Moreover, the leftmost $m-k$ unused inputs in $in$ can be determined in $O(m)$ time by examining the leftmost $m$ inputs of $in$.

Effectively, the first {\em for} loop assigns each active input in the slim section to the output with the same index value, while also inserting each unused output $z_{i}$ into a list $L$ of $m$ elements as it checks if input $y_i, 1\le i\le m$ is active.
The second {\em for} loop then assigns the active inputs from left to right in the fat section to the unused outputs, i.e., those that are inserted into $L$ from top to bottom.

The time complexity of Algorithm~\ref{alg:classicalFull} is easily seen to be $O(n)$ steps, given that (i) each iteration in the first loop involves checking if a bit in the list $in$ is `1', pairing an input in the slim-section with an output or inserting a value into a list of $m$ elements, and (ii) each iteration in the second loop involves checking a bit, pairing an input in the fat-section with an output, and removing an output from the list $L$.
Checking if $in[i] == 1$ clearly takes $O(1)$ time.
It is further assumed that pairing an input and output as well as inserting or removing a value in and out of a list also takes $O(1)$ time.
This is a reasonable assumption, considering that all three operations involve no more than a basic memory read or write operation.

\vspace{-14pt}

\subsection{Bounded Capacity Fat-Slim Concentration Routing}

\label{subsection:classicalBounded}
\vspace{-2pt}\noindent

Algorithm~\ref{alg:classicalSparse} extends the main idea of Algorithm~\ref{alg:classicalFull} to routing active inputs in a bounded capacity fat-and-slim concentrator\footnote{Even though we refer to this construction as a bounded capacity, fat-and-slim concentrator, the fat section on the right is not completely filled with crosspoints as in the case of full-capacity fat-and-slim concentrator.}.
As in Algorithm~\ref{alg:classicalFull}, the inputs are divided into fat and slim sections, but this time, the slim section is placed on the left and the pseudo-fat section is placed on the right in Figure~\ref{fig2}.
An $m$-bit array, named {\em out} is added to mark the outputs.
The first {\em for} loop pairs the active inputs in the slim section with outputs whose indices coincide with those of the active inputs, while marking those outputs by entering `0's into {\em out} in their index positions.
The second {\em for} loop finds the active inputs in the pseudo-fat section on the right.
The last nested loop pairs the active inputs in the pseudo-fat section on the right by searching for an available output from among the set of outputs to which each active is connected by a crosspoint.

Algorithm~\ref{alg:classicalSparse} has an execution time of $O(n)=O(m)$.
\begin{enumerate}
    \item Lines \ref{cb:slimBegin}-\ref{cb:slimEnd} take $O(m)$ time as they involve steps to check if $in[i] = 1,$ pair $x_i$ with $z_i,$ and clear a bit in the $out$ array.
    \item Lines \ref{cb:fatSearchBegin}-\ref{cb:fatSearchEnd} take $O(q)$ time as they involve checking $q$ array elements and adding them to a list.
    \item Lines \ref{cb:fatBegin}-\ref{cb:fatEnd} consist of a nested loop that takes $O(c)$ time.
    To see this, let $c',0\le c'\le c,$ be the number of active inputs found in lines \ref{cb:slimBegin}-\ref{cb:slimEnd}.
    In line \ref{cb:break}, we exit the interior {\em for} loop.
    This happens when an available output is found for that input.
    The number of unsuccessful checks for an available output can at most be $c'$ and successful checks for such an output can at most be $c$.
    Since the number of elements in the list $L$ is also bounded by $c$, this nested loop takes $O(c)$ time.
\end{enumerate}

\begin{algorithm}
\caption{Classical Bounded Capacity F-S Concentration}
\label{alg:classicalSparse}
\small
\begin{algorithmic}[1]
\Function{Classical Bounded F-S Route}{$in,n,m,q$}\newline
//$in$: $n$-bit array that marks up to $c$ active inputs.\newline 
//$n$: number of inputs.\newline 
//$m$: number of outputs.\newline
//$q$: the width of the fat section.\newline
//$c = \lfloor m/q\rfloor$: capacity.\newline
//$out$: $m$-bit array that marks available outputs (initialized to all 1's).
\For{$i \gets 1:n-q$} //slim section \label{cb:slimBegin}
\If {$in[i]==1$}
\State$pair (x_i, z_i);$ $out[i]\gets 0;$
\EndIf
\EndFor\label{cb:slimEnd}

\State $L \gets list()$ //pseudo-fat section \label{cb:fatSearchBegin}

\For{$i \gets 1:q$} 
\If {$in[i+m]==1$}
\State $L.insert(i);$
\EndIf
\EndFor \label{cb:fatSearchEnd}

\While{$i\gets L.remove()$} \label{cb:fatBegin}
\For{$j \gets 0:c-1$}
\If{$out[i+qj]==1$}
\State$pair (y_i,z_{i+qj});$
\State \textbf{break} \label{cb:break}
\EndIf
\EndFor
\EndWhile \label{cb:fatEnd}

\EndFunction 
\end{algorithmic}
\end{algorithm}
\vspace{-5pt}\noindent

\vspace{-11pt}
\subsection{Regular Fat-Slim Concentration Routing}
\label{subsection:classicalRegular}

\vspace{-2pt}\noindent
Our third classical algorithm restates the one given in~\cite{guo96} with a tighter time complexity analysis on some steps.
This time, we have a sparse crossbar construction with $n = pm$ inputs and $m$ outputs as described in Section~\ref{section:concentrators}, where we assume that $p$ divides $m$.
A routing request with $k$ active inputs is first completed to a request with $m$ inputs by combining the first unused $m-k$ inputs in $I_1$ with the given $k$ active inputs.
This is always possible since if all $k \le m$ active inputs belong to $I_1$ then its remaining $m-k$ inputs can be combined with the $k$ active inputs to obtain an $m$-request.
On the other hand, if only some $k' < k \le m$ of the $k$ active inputs belong to $I_1$ then $I_1$ must have $m-k' > m-k$ unused inputs, the first $m-k$ of which can be combined with the given $k$ active inputs to obtain an $m$-request.
Moreover, the first $m-k$ unused inputs in $I_1$ can be determined in $O(m)$ time by examining all $m$ inputs in $I_1$ in the worst case\footnote{The selection of $I_1$ is arbitrary and simplifies our description for the completion of a $k$-assignment to an $m$-assignment.
The algorithm will work regardless of which set of inputs selected.}.
Therefore, we will assume that a $k$-request is completed to an $m$-request and specified by an array of $n$-bits named $in$, in which 0 and 1 bits represent the absence and presence of an active input respectively.
At the beginning of the algorithm, the locations of these active inputs are searched and active inputs in $I_j = \{x_{j,1},x_{j,2},\ldots,x_{j,m}\}$ are placed in a linked list $R_j, 1\le i\le p$.

\vspace{-2pt}
Algorithm~\ref{alg:regularSub} routes an $m$-request by considering two distinct cases:
(a) One of the $R_j$'s has more than $m- m/p$ active inputs.
(b) None of $R_j$'s has more than $m-m/p$ active inputs.
That there can be no other case for all $p \ge 2$ is shown as follows.
Suppose there exists $R_j$ with more than $m - m/p$ active inputs, for some $j$.
Then there remain less than $m-(m-m/p) = m/p$ active inputs in the union of all the remaining $p-1$ sets of $m$ inputs.
Therefore, another $m$-bit array, i.e., $R_{j'}$ for some $j'\ne j, 1\le j'\le p$ with more than $m - m/p$ active inputs exists only if $m/p > m - m/p,$ which implies $p < 2$ or $p = 1,$ contradicting our assumption that $p\ge 2$ as stated in Section~\ref{section:concentrators}.

Now, continuing with Algorithm~\ref{alg:regularSub}, lines 1 through 8 construct the linked lists in $O(n)$ time, assuming that we have $m$ active inputs as described above and determine the number of entries in each linked list by incrementing a size field each time an active input is found.
Next, we see that the {\em for} statement in line~\ref{step:1} is iterated at most $p$ times, which occurs either when none of $R_j, 1\le j\le p$ has more than $m-m/p$ active inputs or $R_p$ does.
During the $j$th iteration, the {\em if} statement in line~\ref{algorith3step:3} checks if  $R_j$ has more than $m-m/p$ 1's.
The size of each set can be queried in $O(1)$ time with the help of the size field  that has been computed in line 5.

\vspace{-3pt}
{\em Case} (a): If there exists such an $R_j$ then line~\ref{step:1.1} pairs all the active inputs in $R_j\cap W_j,$ if any, and the outputs in $U_j$ to which those active inputs are connected by crosspoints.
Line~\ref{step:1.2} then pairs as many active inputs in $R_i, 1\le i\ne j\le p$ as possible, if any, with unused outputs in $U_j$.
At this point, any active remaining inputs in $R_i, 1\le i\ne j\le p$, are paired with unused outputs in $U_l, l\ne j$ in lines~\ref{cr:vj1} and \ref{cr:vj2}.
This pairing is implemented in such a way that one of the $U_i, 1\le i\ne j$ is arbitrarily fixed ($U_{j-1}$ in the algorithm) to start the pairing.
It is possible that $U_{j-1}$ may not even have an active input, but line 13 serves to initiate pairing of the active inputs in the remaining subsets of inputs other than $I_j$.
We note that the number of active inputs in $I_{j-1}$ is less than $m/p=|U_{j+1}|$ and a full crossbar connection exists between $I_{j-1}$ and $U_{j+1}$.
Similarly, in line \ref{cr:vj2}, it is always possible to pair the number of active inputs in $\!\!\!\!\!\bigcup\limits_{\substack{i\ne j,  i\ne j-1}}\!\!\!\!\!\! I_k$ with the outputs in $U_{j-1}$ as $U_{j-1}$ has  $m/p$ available outputs, which are more numerous than the active inputs in $\!\!\!\!\!\bigcup\limits_{\substack{i\ne j,  i\ne j-1}}\!\!\!\!\!\! I_k$, and a full crossbar connection exists between the two sets.
Here, the indexing is cyclical, and it is assumed that $j$, $j-1,$ and $j+1$ are distinct from each other and $p\ge 3$.
If $p = 2$ then line 14 is not needed.
Finally, any remaining inputs in $R_j$ are paired with the remaining unused outputs in $V_j$ in line~\ref{step:1.4} as there is a full crossbar connection between the inputs in $I_j$ and outputs in $V_j, 1\le j\le p$.

\begin{algorithm}
\caption{Classical Regular F-S Concentration}
\label{alg:regularSub}
\label{alg:classicalRegular}
\small
\begin{algorithmic}[1]
\Function{Classical Regular F-S Route}{$in,p,m$}\newline
//$in$: $n$-bit array that marks up to $m$ active inputs.\newline 
//$n=m*p$: number of inputs.\newline 
//$R_j, 1\le j\le p$:  linked lists of active inputs.\newline 
//$m$: number of outputs.\newline
//The size fields of $R_j, 1\le j\le p$ are cleared.
\For{$j\gets 1:p$}
\label{cr:searchStart}
\For{$i\gets 1:m$} //find active elements
\If{$in[(j-1)m + i]==1$}
\State {$R_j.insert(i); R_j.size \scriptstyle{+\!+}$}
\EndIf
\EndFor
\EndFor \label{cr:searchEnd}

\For{$j\gets1:p$}\label{step:1}\label{cr:routingStart}
\If{$R_j.size>m-m/p$} //case (a)\label{algorith3step:3}\label{cr:diagoanlStart}
\State Pair inputs in $R_j\cap W_j$ with outputs in $U_j;$\label{step:1.1}\label{cr:ujj}
\State Pair inputs in $\!\!\underset{i\ne j} {\cup} R_i$ with unpaired outputs in $U_j;$ \label{step:1.2}\label{cr:uji}
\State Pair unpaired inputs in $R_{j-1}$ with outputs in $U_{j+1};$\label{cr:vj1}
\State Pair unpaired inputs in $\!\!\underset{i\ne j, i\ne j-1} {\cup}\!\! R_i$ with outputs in  $U_{j-1};$\label{step:1.3}\label{cr:vj2}
\State Pair unpaired inputs in $R_j$ with unpaired outputs in $V_j;$ \label{step:1.4}\label{cr:vj3}
\State \textbf{return}\label{cr:diagonalEnd}

\Else $\,$//case (b)\label{cr:nondiagonalStart}
\State $a[j]\gets(m/p\le |R_j|);$ \label{step:2.1}\label{step:10}\label{cr:computeA}
//$p$-bit array for reindexing\label{step:13}
\EndIf
\EndFor
\State $r\gets$ prefixSum$(a);$ \label{cr:prefixSumR}  $s\gets $ prefixSum$(\neg a);$\label{step:14}\label{cr:prefixSumS}
\State $d[j]\gets a[j] r[j]+(\neg a[j])(s[j]+r[p]); {1\le j\le p;}$ \label{step:15}\label{cr:newIndices}
//new indices
\State $R'_{d[j]}\gets R_j$\label{cr:reindexR}; $U'_{d[j]}\gets U_j, 1\le j\le p;$ \label{cr:reindexU} \label{step:2.2}
\For{$j\gets 2:r[p]$}\label{step:4}\label{step:16}
\State Take first $m/p$ elements from $R'_j$ and place them in $P'_j;$\label{step:22}
\State Pair inputs in  $P'_j$ with outputs in $U'_{j-1};$ \label{step:4.1}
\State $Q'_j\gets R'_j\backslash P'_j;$\label{step:4.2}
\EndFor
\State Pair inputs in $R'_{r[p]+1},R'_{r[p]+2},...,R'_{p}, R'_1,Q_2,Q_3,...,Q_{r[p]}$  \label{step:5}
\State with outputs in $U'_{r[p]},$ $U'_{r[p]+1},...,U'_{p};$\label{cr:routingEnd}
\EndFunction
\end{algorithmic}
\end{algorithm} 

\begin{figure}
\centerline{\includegraphics[width=\linewidth]{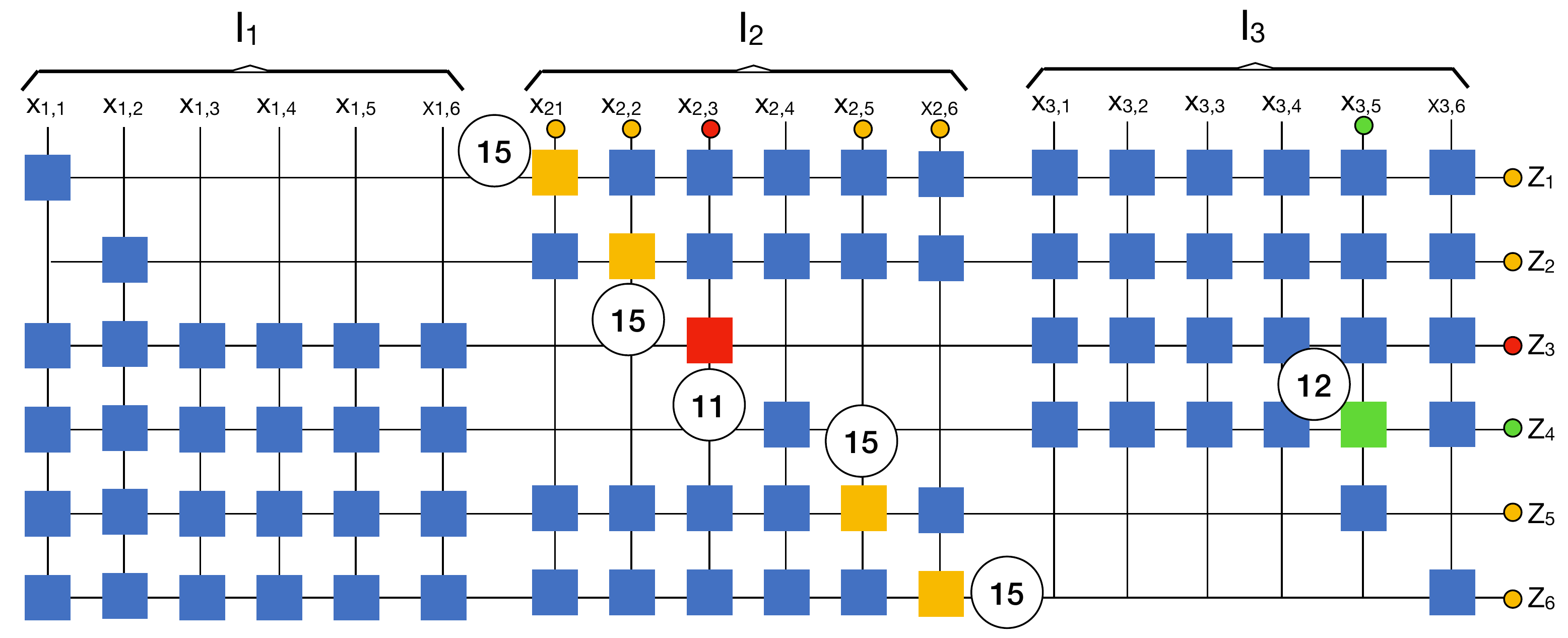}}
\vspace{-6pt}
\caption{Routing in a regular (18,6)-fat-and-slim concentrator.}
\label{fig4}
\end{figure}

\vspace{-2pt}
These steps are illustrated in Figure~\ref{fig4}, where the circled numbers identify the steps in the algorithm.
In this example, lines~\ref{cr:vj1} and \ref{cr:vj2} do not result in any pairing.
As another example, let $m= 12, p = 3 ,$ $R_1 = \{1,2,3,5,6,7,8,9,10\}$, $R_2 = \{3,8\}$, $R_3 =\{1\}$.
We have $U_1 = \{z_1, z_2, z_3, z_4\}, U_2 = \{z_5, z_6, z_7, z_8\}, U_3 =\{z_9, z_{10}, z_{11}, z_{12}\},$ and $|R_1|  = 9 > 8 = m-m/p$.
Thus, $x_{1,1}, x_{1,2}, x_{1,3}$ are paired with $z_1, z_2, z_3$ in $U_1$ in line~\ref{step:1.1}, $x_{2,3}$ is paired with $z_4$ in $U_1$ in line~\ref{step:1.2}, $x_{3,1}$ in $I_3$ is paired with output $z_5$ in $U_2$ and the remaining active input $x_{2,7}$ in $I_2$ is paired with output $z_9$ in $U_2$ in lines \ref{cr:vj1} and \ref{cr:vj2}.
Finally, $x_{1,5}, x_{1,6}, x_{1,7}, x_{1,8}, x_{1,9}, x_{1,10}$ are paired with the remaining outputs all of which belong to $V_1$.
We established that any routing request in case (a) can always be realized in lines~\ref{step:1.1} through~\ref{step:1.4}.
Now suppose that the condition $|R[j]| > m-m/p$ fails for all $j, 1\le j\le p$.
The remaining part of the algorithm handles case (b) as described next.

\vspace{-3pt}
{\em Case} (b): First, in line 17, we initiate a process to identify the subsets of inputs with at most $m/p$ active inputs.
This is done to reindex the sets of inputs $I_j, 1\le j\le p$ so that those that have between $m/p$ and $m-m/p$ active inputs are given lower index values.
Effectively, the $R_j$'s and the corresponding $U_j$'s are implicitly sorted in descending order, based on whether they contain more than the threshold of $m/p$ active inputs.
To facilitate this, we follow the approach in~\cite{guo96} and form a bit-array of $p$ elements, $a$ in line~\ref{step:10} such that $a[j] = 1$ if and only if $m/p \le |R_j|\le m- m/p, 1\le j\le p$.
This array is used to split $R_j$ into two groups, those for which $m/p \le |R_j|\le m- m/p$ and those for which $|R_j| < m/p$.
This splitting is carried out by computing the prefix sums of the bits in $a$ and $\neg a$ into two $p$-element arrays $r$ and $s$ in line~\ref{cr:prefixSumR}, where $r$ ranks  those $R_j$ for which $m/p \le |R_j|\le m- m/p,$ and $s$ ranks those $R_j$ for which $|R_j| < m/p$.
The ranks index the active inputs in each group separately.
The ranks are threaded together in line~\ref{step:15} to obtain a $p$-element array, $d$, which represents a permutation of the indices of $R_j$.
This step essentially involves selecting one of $r[j]$ or $r[p] + s[j]$ into $d[j]$ and is used to compute the indices of $R_j$'s.
It is not difficult to see that $d$ is a permutation of the indices of $R_j$'s and applying $d$ to the indices of $R_j$'s and $U_j$'s amounts to renaming them so that $R_j$ becomes $R_{d[j]}$ and $U_j$ becomes $U_{d[j]}, 1\le j\le p$.
For clarity, we replace $R_{d[j]}$ and $U_j$ by  $R'_{d[j]}$ and $U'_j$ in the algorithm.

\vspace{-3pt}
  As an example, let $m = 20, p = 5, $ and suppose that $I_1, I_2, I_3, I_4, I_5$ have  $3, 4, 6, 5, 2$ active inputs, respectively.
Given that $m/p = 5, m-m/p = 20-20/5 = 16,$  $ |I_1|, |I_2|,|I_5| < m/p$ and $m/p <  |I_3|, |I_4| \le m-m/p$ so that $a = [0, 0, 1, 1, 0], \neg a = [1, 1, 0, 0, 1],$ $r = [0, 0, 1, 2, 2], s = [1, 2, 2, 2, 3], d = ar + (\neg a) ( r[p] + s ) = [0, 0, 1, 2, 0] + [3, 4, 0, 0, 5]  = [3, 4, 1, 2, 5]$.
 Thus, $R'_3\leftarrow R_1, U'_3\leftarrow U_1, R'_4\leftarrow R_2, U'_4\leftarrow U_2,  R'_1\leftarrow R_3,  U'_1\leftarrow U_3,  R'_2\leftarrow R_4, U'_2\leftarrow U_4, R'_5\leftarrow R_5, U'_5\leftarrow U_5$.
This gives a permuted set of registers in descending cardinalities, i.e., $R'_3, R'_4, R'_1, R'_2, R'_5$ and $U'_3, U'_4, U'_1, U'_2, U'_5$.

\vspace{-2pt}
Once this sorting process is completed in line~\ref{cr:reindexU}, we  select the first $m/p$ active inputs in each of the $r[p]$ sets in which the number of active inputs lies between $m/p$ and $m-m/p-1,$ and store all, but the first $m/p$ active inputs into $P'_j,2\le j\le r[p]$ in line~\ref{step:22}.
These active inputs are paired with sets of $m/p$ outputs in $U'_{j-1}, 2\le j\le r[p]$ in line~\ref{step:4.1} using the full crossbar connection between them.
This will complete the pairing of $(r[p]-1)m/p$ active inputs and $m-(r[p]-1)m/p = (p-r[p]+1)m/p$ active inputs remain.

The remaining active inputs in $R'_j$ are saved into $Q'_j, 2\le j\le r[p]$ in line~\ref{step:4.2}.
Finally, the active inputs in $Q_j$'s are concatenated with the active inputs in $\!R_{r[p]+1},R_{r[p]+2},...,R_{[p]},R_1$ which are then paired together with the outputs in sets $U'_{r[p]+1},U'_{r[p]+2},...,U'_{p}$ in line~\ref{cr:routingEnd}.
Note that the number of active inputs in $R'_{r[p]+1},R'_{r[p]+2},...,R'_{p}, R'_1,Q_2,Q_3,...,Q_{r[p]}$ is given by $(p-r[p]+1)m/p$ and it must match the number of active outputs in in $U'_{r[p]},$ $U'_{r[p]+1},...,U'_{p}$ after the pairing in line~\ref{step:4.1}.

Algorithm~\ref{alg:regularSub} has a time complexity of $O(n)$ as shown below.

\vspace{-2pt}
\begin{enumerate}
\item Lines \ref{cr:searchStart}-\ref{cr:searchEnd}:
This is a linear search of active inputs out of $n$ inputs, and therefore has a time complexity of $O(n)$.

\item Lines \ref{cr:routingStart}-\ref{cr:routingEnd}:
By the time we get to these lines, we already know where the active inputs are located because $\!R_j$'s  consist only of active inputs.
At this point the only task left to do is to route these active inputs.
We route in two different ways as described in cases (a) and (b) in the algorithm.
Both take $O(m)$ time as described below.

\vspace{-4pt}
\begin{enumerate}
 \item Lines~\ref{cr:routingStart} and~\ref{cr:diagoanlStart}: We see that the {\em for} statement here is iterated at most $p$ times, which occurs either when none of $R_j, 1\le j\le p$ has more than $m-m/p$ active inputs or $R_p$ does.
 During the $j$th iteration, the {\em if} statement in line~\ref{algorith3step:3} checks if $R_j$ has more than $m-m/p$ elements.
 Since we can query the size of sets in $O(1)$ time and comparison of values in the if statement can also be completed in $O(1)$ time, this line has a time complexity of $O(p)$. \vspace{-2pt}
 \item Lines \ref{step:1.1}-\ref{cr:diagonalEnd}:
 In case (a), there are more active inputs than non-diagonals can handle alone, and therefore diagonal crosspoints must be used.
These lines execute at most once and take $O(m)$ time as explained below.

\vspace{-4pt}
 \begin{enumerate}
     \item Line \ref{cr:ujj} can be done by iterating through $R_j$ and checking if the current element is in $W_j$.
Since $W_j$ is an interval where endpoints are fixed and known, the membership decision amounts to the comparison of the current element against the two endpoints, and this takes $O(1)$ time in our classical bit-parallel processor model.
    Moreover, the number of pairing operations is clearly bounded by $|W_j| = |U_j| = m/p,$ and pairing an active input with the corresponding output in $U_j$ takes $O(1)$ time.
Therefore, this line can be done in $O({\rm max }(m,m/p)) = O(m)$ time, the size limit of $R_j$.
    
\vspace{1pt}  
     \item Line \ref{cr:uji}: Let  $\hat{R}  = \underset{i\ne j} {\cup} R_i$.
 Given that all inputs in  $\hat{R}$  are connected to outputs in $U_j$ by a full crossbar connection, the active inputs in $\hat{R}$ can be paired with unused outputs in $U_j$ by scanning $\hat{R}$ from left to right and $U_j$ from top to bottom until all unused outputs in $U_j$ are paired.
As we represent $R_j, 1\le j \le p$ by linked lists and since there are $m$ active inputs at the most, the pairing can be completed in $O(m$) time in this case. \vspace{-2pt}
     \item In line \ref{cr:vj1} we pair two sides of a full crossbar connection, i.e., inputs in $R_{j-1}$ and outputs in $U_{j+1}$.
We loop through these sets, removing one element from each to build a complete matching.
Given that $|R_{j-1}| \le m, 2\le j \le m+1,$ and $|U_{j+1}| = m/p,$  this line has a time complexity of $O(m)$.
     \vspace{-2pt}
     \item In line 14, we proceed as in line~\ref{cr:uji}, and note that  $ |\!\!\underset{i\ne j, i\ne j-1}{\cup}\!\!\!\! R_i|\!\le\! m$.
Therefore this line has a time complexity of  $O(m)$.
\vspace{-2pt}
     \item Line~\ref{cr:vj3} involves connecting the remaining active inputs in $I_j$ and the unused outputs in $V_j$.
     It is not difficult to verify that this step can be completed in $O(m - m/p)$ time as there is a full crossbar connection between the inputs in $R_j$ and $V_j, 1\le j\le p$.
     It follows that the time complexity of the {\em for} loop is $O(m - m/p)$.
 \end{enumerate}
 \vspace{-3pt}
 \item Lines \ref{cr:nondiagonalStart}-\ref{cr:routingEnd}:
 These lines are carried out in case (b) only.
In this case, non-diagonal crosspoints can handle the assignment without diagonal crosspoints.
Therefore, only the fat sections are used in the routing and these lines can be done in $O(m)$ time as explained below.

\vspace{-2pt}
 \begin{enumerate}
     \item Line \ref{cr:computeA} involves querying size of $R_j$ and comparing it to $m/p$, which can be completed in $O(1)$ time per each iteration of the {\em for} loop and in $O(p)$ time for all iterations. \vspace{-1pt}
     \item Line \ref{cr:prefixSumR}  has prefix sum of $p$ elements that can be done in $O(p)$ time by an iterative summation of $p$ bits.\vspace{-1pt}
     \item Line \ref{cr:newIndices} involves an addition of two $O(\lg p)$-bit numbers and a $2\times 1$ multiplexer, both of which take $O(1)$ time and so the loop in this line takes  $O(p)$ time. \vspace{-1pt}
     \item Line \ref{cr:reindexR}  involves reindexing of $R_j$'s, which clearly takes $O(p)$ time. \vspace{-1pt}
     \item Lines \ref{step:16}-\ref{step:4.2} can be carried out in $O(m/p)$ time per each set of $m/p$ inputs and in $O(r[p]\times m/p) = O(m)$ time for all $r[p]$ sets of $m/p$ inputs. \vspace{-1pt}
     \item Lines \ref{step:5}-\ref{cr:routingEnd} can be done by pairing the elements between the two sets described in these lines one by one  in $O(m)$ time.
 \end{enumerate}
  
\end{enumerate}
\end{enumerate}

\vspace{-7pt}

\begin{remark}
\label{remark1}
{\rm Algorithm~\ref{alg:classicalFull}}, {\rm Algorithm~\ref{alg:classicalSparse} and {\rm Algorithm~\ref{alg:classicalRegular}} are all optimal in terms of order of execution time as it takes $\Omega(n)$ time to read a concentration assignment.\qed}
\end{remark}

\vspace{-16pt}
\section{Quantum Routing-Grover's Search}
\label{section:quantumRouting}

\vspace{-4pt}\noindent
Now that we have established that classical routing algorithms all take $O(n)$ time for full and bounded capacity fat-and-slim concentrators described in Section \ref{section:concentrators}, we turn our attention to quantum routing for them.
Since both these concentrators require searching of active inputs out of all inputs, Grover's search \cite{Gro96} provides a possible approach to identify such inputs faster in the quantum domain.
Grover's search approach has been followed to speed up various matching and network flow problems in \cite{AS06} and \cite{Dor09}.
These two papers were among the inspirations for our approach.
We will use the version of Grover's search analyzed in Theorem 3 of \cite{BBHT98} and refer to it as {\em GroverSearch} in this paper.
We note that {\em GroverSearch} is the only quantum step in the algorithms to be presented in this section and rest of the paper.
The time complexity of {\em GroverSearch} is known to be $O(\sqrt{n/k})$ to find one of $k$ marked items out of $n$ possible items.
Thus, it takes $O(\sqrt{nk})$ time to discover all $k$ items by unmarking discovered items one at a time.
This is a widely used method even though the original source is unknown.
The proof can be found in Lemma 4.1 of \cite{CK12}.

\vspace{-5pt}
\begin{remark}
\label{remark:monteCarlo}
{\rm As in \cite{AS06} and \cite{Dor09}, we use Monte Carlo amplification to bound the total error of the entire algorithm resulting in an increase in run-time that is a logarithmic factor in the number of calls to the quantum subroutine, i.e., $\ln k$.
Therefore, the time to discover all marked inputs is $O(\sqrt{nk}\ln{k})$.} \qed
\end{remark}

\vspace{-5pt}
\begin{algorithm}
\caption{Quantum Full F-S Concentration}
\label{alg:quantumFull}
\small
\begin{algorithmic}[1]
\vspace{2pt}
\Function{Quantum Full FS Route}{$in,n,m$}\newline
//$in$: $n$-bit array that marks up to $m$ active inputs\newline 
//$n$: number of inputs.\newline 
//$m$: number of outputs.
\State $L \gets list()$

\For{$i \gets 1:m$} //slim section
\If {$in[m-n+i] == 1$}
\State$pair (y_i, z_i);$
\Else \vspace{-1pt}
\State $L.insert(z_i);$\vspace{1pt}
\EndIf
\EndFor
\While {$i\gets GroverSearch(in[1:n-m])$}//fat section
\State$z=L.remove(); pair(x_i, z);$
\State$in[i]=0$ \vspace{1pt}
\EndWhile \vspace{1pt}
\EndFunction

\end{algorithmic}
\end{algorithm}

\vspace{-20pt}
\subsection{Quantum Full-Capacity F-S Concentration}
\label{subsection:quantumFull}

\vspace{-3pt}\noindent
The pseudo-code of our routing algorithm for the full-capacity fat-and-slim concentrator is given in Algorithm~\ref{alg:quantumFull}.
As in the classical case, this algorithm consists of two parts.
In the first part, we route all the active inputs to corresponding outputs in the slim section.
We also make a list of idle inputs in this part to identify the corresponding outputs that are not used by the inputs in the slim section.
Such outputs are then paired with the active inputs that are found by {\em GroverSearch} in the second part of the algorithm.
This part of the algorithm is classical and deterministic.
The second part of the algorithm provides us with quantum speedup.
We use {\em GroverSearch} to locate the active inputs in the fat part of the concentrator.
Once an active input is found, we pair it with one of the idle outputs that were found in the first part and unmark that input for another round of {\em GroverSearch}.

The time complexity of Algorithm~\ref{alg:quantumFull} is computed as follows.
First note that the slim part of the algorithm is classical and deterministic.
Since we assume that each query of the array $in$ requires a constant time, identifying the active inputs, assigning them to their respective outputs, and also identifying the unused outputs in the slim part takes $O(m)$ time.
Now, let $k$ be the number of active inputs in the fat part.
It is obvious that $k$ is upper bounded by $m$.
Hence by Remark \ref{remark:monteCarlo}, we find that the total time required for the fat part is $O(\sqrt{m(n-m)}\ln m)$.
Therefore the total time complexity of concentration in a full-capacity, fat-and-slim $(n,m)$-concentrator is $O(m+\sqrt{k(n-m)}\ln m) = O(m+\sqrt{nm}\ln m)$, and hence, given that $n \ge m, $ the total time complexity is  $O(\sqrt{nm}\ln m),$ using a quantum processor with $O(\ln n)$ qubits.
For comparison, any classical-bit processor routing algorithm for a full capacity, fat-and-slim $(n,m)$-concentrator takes $\Theta(n)$ time as we established in the earlier section (See Remark~\ref{remark1}).
When $n=\Theta(m\ln^{2} m)$, we have $O(\sqrt{nm}\ln m)=O(n)$.
Therefore, our algorithm performs better in quantum domain than any known classical algorithm if $m\ln^2{m}=o(n)$.
One such value of $m$ is clearly $\Theta(\sqrt{n})$.
In general, it can easily be shown that $m = n^\mu,$ satisfies $m\ln^{2} m = o(n)$, for any $\mu, 0 < \mu < 1,$ and the time complexity of our algorithm will increase less than linearly with $n$ for such values of $m$.

\vspace{-10pt}
\subsection{Quantum Bounded-Capacity F-S Concentration}

\vspace{-2pt}\noindent
The time complexity of the classical routing algorithm for a bounded capacity fat-and-slim concentrator can also be reduced using quantum search as well, as described in Algorithm~\ref{alg:quantumSparse}.
The slim section of the algorithm finds at most $c$ marked elements out of at most $m$ total elements since $n-q\le m$.
Similarly, the fat section of the algorithm also finds $c$ marked elements out of at most $m$ total elements since $q\le m$.
The total time complexity of both parts is $O(\sqrt{mc}\ln{c})$.
Lines \ref{qb:fatBegin}-\ref{qb:fatEnd} is the same as \ref{cb:fatBegin}-\ref{cb:fatEnd} of Algorithm \ref{alg:classicalSparse}.
Therefore, the total time complexity is also $O(\sqrt{mc}\ln{c})=O(\sqrt{nc}\ln{c})$.
Once again, this algorithm perform better than it's classical counterpart when $c\ln^2{c}=o(n)$.

\vspace{-5pt}
\begin{algorithm}
\caption{Quantum Bounded FS Concentration}
\label{alg:quantumSparse}
\small
\begin{algorithmic}[1]
\Function{Quantum Sparse FS Route}{$in,n,m,q$}\newline
//$in$: $n$-bit array that marks up to $c$ active inputs.\newline 
//$n$: number of inputs.\newline 
//$m$: number of outputs.\newline
//$q$: the width of the fat section.\newline
//$c = \lfloor m/q\rfloor$: capacity.\newline
//$out$: $m$-bit array that marks available outputs (initialized to all 1's).
\While {$i\gets GroverSearch(in[1:n-q])$} //slim section
\State$pair(x_i,z_i);$
\State$in[i]\gets0;$
\State$out[i]\gets0;$
\EndWhile

\State $L \gets list();$ //pseudo-fat section

\While {$i\gets GroverSearch(in[n-q+1:n])$} //slim section
\State $L.insert(i);$
\State$in[i]\gets0;$
\EndWhile

\While{$i\gets L.remove()$}\label{qb:fatBegin}
\For{$j \gets 0:c-1$}
\If{$out[i+qj]==1$}
\State$pair (y_i,z_{i+qj});$
\State \textbf{break}
\EndIf
\EndFor
\EndWhile\label{qb:fatEnd}

\EndFunction
\end{algorithmic}
\end{algorithm}

\begin{remark}
{\rm In full capacity F-S concentrator, active inputs for the dense part are located with Grover's Search while for bounded capacity FS concentrator, all active inputs are located with Grover's Search. \qed}
\end{remark}

\vspace{-15pt}
\subsection{Quantum Regular FS Concentration}

\noindent
Our last quantum algorithm handles concentration assignments for regular fat-and-slim concentrators.
It is obtained by replacing the search for active inputs by {\em GroverSearch}.
From the analysis in subsection \ref{subsection:classicalRegular} we know that once the active inputs are found, the routing takes $O(m)$. Just like full capacity fat-and-slim concentrator, the search as well as the total time takes $O(\sqrt{nm}\ln{m})$.
Therefore the analysis done in subsection \ref{subsection:quantumFull} also applies here.
\begin{algorithm}
\caption{Quantum Regular F-S Concentration}
\label{alg:quantumRegular}
\small
\begin{algorithmic}[1]
\Function{Quantum Regular F-S Route}{$in,p,m$}\newline
//$in$: $n$-bit array that marks up to $m$ active inputs.\newline 
//$n=m*p$: number of inputs.\newline 
//$R_j, 1\le j\le p$:  linked lists of active inputs.\newline 
//$m$: number of outputs.\newline
//The size fields of $R_j, 1\le j\le p$ are cleared.
\While {$(i,j)\gets GroverSearch(in)$}
\State {$R_j.insert(i); R_j.size \scriptstyle{+\!+}$}
\EndWhile

\For{$j\gets1:p$}
\If{$R_j.size>m-m/p$} //case (a)
\State Pair inputs in $R_j\cap W_j$ with outputs in $U_j;$
\State Pair inputs in $\!\!\underset{i\ne j} {\cup} R_i$ with unpaired outputs in $U_j;$ 
\State Pair unpaired inputs in $R_{j-1}$ with outputs in $U_{j+1};$
\State Pair unpaired inputs in $\!\!\underset{i\ne j, i\ne j-1} {\cup}\!\! R_i$ with outputs in  $U_{j-1};$
\State Pair unpaired inputs in $R_j$ with unpaired outputs in $V_j;$ 
\State \textbf{return}

\Else $\,$//case (b)
\State $a[j]\gets(m/p\le |R_j|);$ 
//$p$-bit array for reindexing
\EndIf
\EndFor
\State $r\gets$ prefixSum$(a);$ 
\State $d[j]\gets a[j] r[j]+(\neg a[j])(s[j]+r[p]); {1\le j\le p;}$ 
//new indices
\State $R'_{d[j]}\gets R_j$
\For{$j\gets 2:r[p]$}
\State Take first $m/p$ elements from $R'_j$ and place them in $P'_j;$
\State Pair inputs in  $P'_j$ with outputs in $U'_{j-1};$ 
\State $Q'_j\gets R'_j\backslash P'_j;$
\EndFor
\State Pair inputs in $R'_{r[p]+1},R'_{r[p]+2},...,R'_{p}, R'_1,Q_2,Q_3,...,Q_{r[p]}$  
\State with outputs in $U'_{r[p]},$ $U'_{r[p]+1},...,U'_{p};$
\vspace{2pt}
\EndFunction
\end{algorithmic}
\end{algorithm} 

\vspace{-5pt}
\section{Conclusions and Future Work}
\label{section:conclusion}

\noindent
We have presented three quantum algorithms, all with smaller time complexity as compared to their classical counterparts.
For full-capacity fat-and-slim concentrator, our quantum algorithm has a time complexity of $O(\sqrt{nm}\log m)$ versus the classical routing algorithm complexity of $O(n)$.
Thus, in this case, our quantum algorithm has a smaller time complexity if 
$m = n^\mu, 0 < \mu < 1$.
For regular fat-and-slim and bounded-capacity fat-and-slim concentrators, similar speed-up formulas apply as established in the paper.
It should be noted in both classical and quantum domains, search process can be speeded up by replicating computational resources.
In the case of a classical search algorithm, using $k$ processors results in a factor of $k$ reduction in time complexity from $O(n)$ to $O(n/k)$.
On the other hand, in the quantum domain, this reduction in time will be limited to a factor $\sqrt{k}$.
Thus, the separation in time complexities between the classical and quantum domains become more pronounced only when $k$ is small.

As for future research, we will investigate the possibility of quantum accelerations for other concentrator architectures and explore other quantum algorithms such as quantum walk in our research.  Another related problem where a similar approach might be helpful is routing on packet switching networks. These and other related problems will be further explored in another paper.

\vspace{-10pt}


\begin{thebibliography}{10}
\providecommand{\url}[1]{#1}
\csname url@samestyle\endcsname
\providecommand{\newblock}{\relax}
\providecommand{\bibinfo}[2]{#2}
\providecommand{\BIBentrySTDinterwordspacing}{\spaceskip=0pt\relax}
\providecommand{\BIBentryALTinterwordstretchfactor}{4}
\providecommand{\BIBentryALTinterwordspacing}{\spaceskip=\fontdimen2\font plus
\BIBentryALTinterwordstretchfactor\fontdimen3\font minus
  \fontdimen4\font\relax}
\providecommand{\BIBforeignlanguage}[2]{{%
\expandafter\ifx\csname l@#1\endcsname\relax
\typeout{** WARNING: IEEEtran.bst: No hyphenation pattern has been}%
\typeout{** loaded for the language `#1'. Using the pattern for}%
\typeout{** the default language instead.}%
\else
\language=\csname l@#1\endcsname
\fi
#2}}
\providecommand{\BIBdecl}{\relax}
\BIBdecl

\bibitem{childs2010quantum}
A.~M. Childs and W.~Van~Dam, ``Quantum algorithms for algebraic problems,''
  \emph{Reviews of Modern Physics}, vol.~82, no.~1, p.~1, 2010.

\bibitem{mosca2008quantum}
M.~Mosca, ``Quantum algorithms,'' \emph{arXiv preprint arXiv:0808.0369}, 2008.

\bibitem{montanaro2016quantum}
A.~Montanaro, ``Quantum algorithms: an overview,'' \emph{npj Quantum
  Information}, vol.~2, no.~1, pp. 1--8, 2016.

\bibitem{AS06}
A.~Ambainis and R.~{\v{S}}palek, ``Quantum algorithms for matching and network
  flows,'' in \emph{Annual Symposium on Theoretical Aspects of Computer
  Science}.\hskip 1em plus 0.5em minus 0.4em\relax Springer, 2006, pp.
  172--183.

\bibitem{ambainis2007quantum}
A.~Ambainis, ``Quantum walk algorithm for element distinctness,'' \emph{SIAM
  Journal on Computing}, vol.~37, no.~1, pp. 210--239, 2007.

\bibitem{Dor09}
S.~D{\"o}rn, ``Quantum algorithms for matching problems.'' \emph{Theory Comput.
  Syst.}, vol.~45, no.~3, pp. 613--628, 2009.

\bibitem{childs2003exponential}
A.~M. Childs, R.~Cleve, E.~Deotto, E.~Farhi, S.~Gutmann, and D.~A. Spielman,
  ``Exponential algorithmic speedup by a quantum walk,'' in \emph{Proceedings
  of the thirty-fifth annual ACM symposium on Theory of computing}, 2003, pp.
  59--68.

\bibitem{pinsker73}
M.~S. Pinsker, ``On the complexity of a concentrator,'' in \emph{7th
  International Telegraffic Conference}, vol.~4.\hskip 1em plus 0.5em minus
  0.4em\relax Citeseer, 1973, pp. 1--318.

\bibitem{pippenger77}
N.~Pippenger, ``Superconcentrators,'' \emph{SIAM Journal on Computing}, vol.~6,
  no.~2, pp. 298--304, 1977.

\bibitem{chu78}
F.~Chung, ``On concentrators, superconcentrators, generalizers, and nonblocking
  networks,'' \emph{The Bell System Technical Journal}, vol.~58, no.~8, pp.
  1765--1777, 1979.

\bibitem{bassalygo81}
L.~Bassalygo, ``Asymptotically optimal switching circuits,'' \emph{Problems of
  Information Transmission}, vol.~17, no.~3, pp. 206--211, 1981.

\bibitem{nakamuraMasson82}
S.~Nakamura and G.~M. Masson, ``Lower bounds on crosspoints in concentrators,''
  \emph{IEEE Transactions on Computers}, no.~12, pp. 1173--1179, 1982.

\bibitem{chienOruc94}
M.~V. Chien and A.~Y. Oru{\c{c}}, ``High performance concentrators and
  superconcentrators using multiplexing schemes,'' \emph{IEEE Transactions on
  Communications}, vol.~42, no.~11, pp. 3045--3050, 1994.

\bibitem{orucHuang96}
A.~Y. Oruc and H.~Huang, ``Crosspoint complexity of sparse crossbar
  concentrators,'' \emph{IEEE Transactions on Information Theory}, vol.~42,
  no.~5, pp. 1466--1471, 1996.

\bibitem{gunduzhanOruc97}
E.~G{\"u}nd{\"u}zhan and A.~Y. Oru{\c{c}}, ``Structure and density of sparse
  crossbar concentrators.'' in \emph{Advances in Switching Networks}.\hskip 1em
  plus 0.5em minus 0.4em\relax Citeseer, 1997, pp. 169--180.

\bibitem{GO98}
W.~Guo and A.~Y. Oru{\c{c}}, ``Regular sparse crossbar concentrators,''
  \emph{IEEE Transactions on Computers}, vol.~47, no.~3, pp. 363--368, 1998.

\bibitem{ratanOruc2003}
R.~Ratan and A.~Oru{\c{c}}, ``Performance evaluation of inputqueued buffered
  sparse-crossbar packet concentrators,'' in \emph{Proc. Conference on
  Information Sciences and Systems CISS}, vol.~3.\hskip 1em plus 0.5em minus
  0.4em\relax Citeseer, 2003.

\bibitem{ratanOruc2010}
R.~Ratan and A.~Y. Oruc, ``Self-routing quantum sparse crossbar packet
  concentrators,'' \emph{IEEE Transactions on Computers}, vol.~60, no.~10, pp.
  1390--1405, 2010.

\bibitem{margulis73}
G.~A. Margulis, ``Explicit constructions of concentrators,'' \emph{Problemy
  Peredachi Informatsii}, vol.~9, no.~4, pp. 71--80, 1973.

\bibitem{gabberGalil78}
O.~Gabber and Z.~Galil, ``Explicit constructions of linear-sized
  superconcentrators,'' \emph{Journal of Computer and System Sciences},
  vol.~22, no.~3, pp. 407--420, 1981.

\bibitem{Alon81}
N.~Alon, ``On the number of subgraphs of prescribed type of graphs with a given
  number of edges,'' \emph{Israel Journal of Mathematics}, vol.~38, no. 1-2,
  pp. 116--130, 1981.

\bibitem{tanner84}
R.~M. Tanner, ``Explicit concentrators from generalized n-gons,'' \emph{SIAM
  Journal on Algebraic Discrete Methods}, vol.~5, no.~3, pp. 287--293, 1984.

\bibitem{jimboMaruoka85}
S.~Jimbo and A.~Maruoka, ``Expanders obtained from affine transformations,'' in
  \emph{Proceedings of the seventeenth annual ACM symposium on Theory of
  computing}, 1985, pp. 88--97.

\bibitem{fins}
A.~Y. Oruc, \emph{Foundations of Interconnection Networks}, 2020.

\bibitem{brent1982regular}
R.~P. Brent and H.~T. Kung, ``A regular layout for parallel adders,''
  \emph{IEEE transactions on Computers}, no.~3, pp. 260--264, 1982.

\bibitem{buvzek1996quantum}
V.~Bu{\v{z}}ek and M.~Hillery, ``Quantum copying: Beyond the no-cloning
  theorem,'' \emph{Physical Review A}, vol.~54, no.~3, p. 1844, 1996.

\bibitem{shor1994algorithms}
P.~W. Shor, ``Algorithms for quantum computation: discrete logarithms and
  factoring,'' in \emph{Proceedings 35th annual symposium on foundations of
  computer science}.\hskip 1em plus 0.5em minus 0.4em\relax Ieee, 1994, pp.
  124--134.

\bibitem{Gro96}
L.~K. Grover, ``A fast quantum mechanical algorithm for database search,'' in
  \emph{Proceedings of the twenty-eighth annual ACM symposium on Theory of
  computing}, 1996, pp. 212--219.

\bibitem{guo96}
W.~Guo and A.~Y. Oruc, ``Explicit construction of bounded capacity sparse
  crossbar concentrators,'' in \emph{Proceedings of the Conference on
  Information Sciences and Systems}, vol.~1.\hskip 1em plus 0.5em minus
  0.4em\relax Department of Electrical Engineering, Johns Hopkins University.,
  1996, p. 233.

\bibitem{BBHT98}
M.~Boyer, G.~Brassard, P.~H{\o}yer, and A.~Tapp, ``Tight bounds on quantum
  searching,'' \emph{Fortschritte der Physik: Progress of Physics}, vol.~46,
  no. 4-5, pp. 493--505, 1998.

\bibitem{CK12}
A.~M. Childs and R.~Kothari, ``Quantum query complexity of minor-closed graph
  properties,'' \emph{SIAM Journal on Computing}, vol.~41, no.~6, pp.
  1426--1450, 2012.

\end{thebibliography}
\end{document}